\def\author{Gary J. Ferland}
\def\letterbody{

What, if any, was Einstein's biggest mistake, the one most affecting our physics today?  
There is a perhaps apocryphal story, recounted  by George Gamow, that he counted his 
cosmological constant as his biggest blunder.  
O’Raifeartaigh and Mitton have recently$^1$ 
summarized the available evidence and conclude that
``We also find it quite plausible that Einstein made such a statement to Gamow in particular.''
In any case, we now know his hypothesized cosmological constant to be correct$^{2,3}$.  
Virginia Trimble has argued that his lifelong rejection of quantum mechanics,
an interesting side-story in the evolution of 20$^{th}$ century physics, is a candidate.  
None of these introduced difficulties in how our physics is done today.

It can be argued that his biggest actual mistake,
one that affects many subfields in physics and chemistry and bewilders students today,
occurred in his classic paper that first derived his $A$ and $B$ coefficients from considerations of the 
blackbody law and statistical equilibrium$^4$.  
The Einstein $A$ is the rate of spontaneous emission of photons 
as an electron moves from an upper to lower energy level and is
universally called his transition probability.
 Energy is conserved during the transition, 
 so the photon energy is equal to the difference in energies between the two levels.  
 Although electronic transitions produce most of the optical / UV / X-ray line emission, 
 other transitions are possible. 
 These include changes in the rotation or vibration of a molecular or the nuclear spin of an atom.  
 The Einstein $A$ applies in all of these cases, so the idea has very broad applicability.
 
 In statistics, a probability is a dimensionless number between 0 and 1.  
 Einstein's $A$ has dimensions s$^{-1}$ and ranges between,
 for example, $\sim 3 \times 10^{-15}$ s$^{-1}$ for the H I 21 cm line in the radio to 
$\sim 3 \times 10^{14}$ s$^{-1}$ for the Fe XXVI K$\alpha$ line in the X-ray.  
 Einstein's  $A$ is clearly not a probability.
 
 In his original 1917 paper, Einstein introduced his $A$ as follows:
 \begin{quote}
Die Wahrscheinlichkeit dW dafut, dass dies im Zeitelement $dt$ wirklich stattfinde, sei
\setcounter{equation}{5}
\begin{equation}
 dW = A_m^n dt 
 \end{equation}
wobei $A_m^n$ eine fur die betrachtete Indexkombination charakteristische Konstante bedeutet.
 \end{quote}

\noindent 
which can be translated as:
 \begin{quote}
Let the probability $dW$ for this to happen during the time interval $dt$, be
\setcounter{equation}{5}
\begin{equation}
 dW = A_m^n dt 
 \end{equation}
where $A_m^n$ is a constant characterizing the index combination under consideration.
 \end{quote}
 The German word ``Wahrscheinlichkeit'' unambiguously translates to the English ``probability''.

The source of the confusion is the definition of $dW$.  
If $dW$ were really the probability over a time interval $dt$, 
then $A$ would be the probability per unit time, 
what we now call the ``transition probability''.  
Actually, $dW$ is the number of photons emitted over the time $dt$.  
The $A$ is actually the rate photons are emitted, the Einstein transition \emph{rate}.  

The term transition ``probability'' creates considerable confusion for students of astrophysical spectroscopy.  
Consider the quantities that enter into the description of how matter emits (section 3.5 of this$^5$ text).  
Collisions can cause transitions within a two-level atom with upper and lower levels $u$ and $l$.  
At the most basic level, collisions between $u$ and $l$ are described in terms of a 
quantum-mechanical cross section $\sigma$ (cm$^2$) which depends on 
the velocity $u$ (cm s$^{-1}$) of the colliding particle.  
This cross section is integrated over a velocity distribution, usually Maxwellian, 
to obtain a rate coefficient $q_{ul}$ with the strange units cm$^3$ s$^{-1}$.  
The rate of collisional transitions $c_{ul}$ is $c_{ul} = q_{ul} n$ (s$^{-1}$)
where $n$ is the density of colliders (cm$^{-3}$). 
The total rate that electrons move from $u$ to $l$ is then $r_{ul} = q_{ul} n + A_{ul}$.  
Knowledgeable students who understand dimensional analysis are bewildered by the concept of 
adding a rate and a probability.  

Although the term ``transition probability'' is an established part of the field$^6$, 
the only part of quantitative spectroscopy which employs something like a probability 
involves many-level systems.  
The  \emph{branching ratio} from a level $u$ is defined as 
$R_{ul} = A_{ul}/\sum_{k<u} A_{uk} $.  
This is the probability that level $u$ will decay by the route $u \rightarrow l$, is dimensionless, 
and ranges between 0 and 1.

$A$ is Einstein's transition rate.

\emph{Acknowledgements}:  Support by NSF (1816537), NASA (ATP 17-ATP17-0141), and STScI (HST-AR- 15018)
is gratefully acknowledged.  

}
\def\department{Physics \& Astronomy}
\def\institute{The University of Kentucky}
\def\address{Lexington}
\def\town{Kentucky}
\def\county{USA}
\def\postcode{40506}
\def\email{gary@uky.edu}
\def\date{2019 March 10}
\def\title{Einstein's biggest mistake?}
\documentclass[12pt]{article}

\begin{document}

\centerline{CORRESPONDENCE}

\bigskip

\centerline{\it To the Editors of `The Observatory'}

\bigskip

\centerline{\it \title}

\bigskip

\indent\letterbody 

\bigskip

\begin{flushright}
{Yours faithfully,\hspace*{2cm}} \\ \sc\author.
\end{flushright}
\footnotesize
\begin{flushleft}
\department, \\ \hspace{.3cm} \institute, \\ \hspace{.6cm} \address,
\\ \hspace{.9cm} \town, \county, \postcode \\ \hspace{1.2cm} [\email]

\bigskip

\date
\end{flushleft}
\normalsize
\centerline{\it References}

1. O'Raifeartaigh, C \& Mitton, S. 2018, Physics in Perspective, Volume 20, Issue 4, pp.318-341

2. Perlmutter, S., et al., 1998, Nature, 391, 51

3. Riess, A.G. et al. 1998, AJ, 116, 1009

4. Einstein, A., 1917, Physikalische Zeitschrift, 18, 121-128

5. Osterbrock, D. E., \& Ferland, G. J. 2006, Astrophysics
of gaseous nebulae and active galactic nuclei, 2nd. ed.
(Sausalito, CA: University Science Books)

6. The NIST online database of Einstein transition rates is accessible at 
\begin{verbatim}
https://physics.nist.gov/PhysRefData/ASD/lines_form.html
\end{verbatim}

\end{document}